\documentclass{elsart}

   \usepackage{graphicx}
   \usepackage{epstopdf}
   \usepackage{amssymb}
   \usepackage{natbib}

\def\teq#1{$\, #1\,$}                         
\def\apj{ApJ}

\def\nat{Nature}
\def\app{Astroparticle Phys.}                   
\def\apss{Astr. Space Sci.}                     
\def\asr{Adv. Space Res.}                       

\def\mnras{{M.N.R.A.S.}}
\def\prl{Phys. Rev. Lett.}                      
\def\prd{Phys. Rev. D}                          
\def\ssr{Space Sci. Rev.}                       
\def\reference{\par \noindent \hangafter=1 \hangindent=0.4 true cm}
\newcommand{\vol}[2]{$\;$\bf #1\rm , #2.}           
             \font\sevenrm=cmr7

          \font\sixrm=cmr6       

\def\machson{{M}_{\hbox{\sixrm S}}}

\def\erg{\varepsilon_\gamma}
\def\taut{\tau_{\hbox{\sixrm T}}}
\def\thetascatt{\theta_{\hbox{\sevenrm scatt}}}
\def\thetaBone{\Theta_{\hbox{\sevenrm Bf1}}}

\def\betaHTone{\beta_{\hbox{\sevenrm 1HT}}}

\begin{document}

\newcommand{\figureoutpdf}[5]{\centerline{}
   \centerline{\hspace{#3in} \includegraphics[width=#2truein]{#1}}
   \vspace{#4truein} \caption{#5} \centerline{} }

\begin{frontmatter}

\title{Using Gamma-Ray Burst Prompt Emission\\ to Probe Relativistic Shock Acceleration}

\author{Matthew G. Baring}
\address{Department of Physics and Astronomy, MS-108,
                      Rice University, P. O. Box 1892, \\
                      Houston, TX 77251-1892, USA}
\ead{baring@rice.edu}

\begin{abstract}
It is widely accepted that the prompt transient signal in the 10 keV --
10 GeV band from gamma-ray bursts (GRBs) arises from multiple shocks
internal to the ultra-relativistic expansion.  The detailed
understanding of the dissipation and accompanying acceleration at these
shocks is a currently topical subject.  This paper explores the
relationship between GRB prompt emission spectra and the electron (or
ion) acceleration properties at the relativistic shocks that pertain to
GRB models. The focus is on the array of possible high-energy power-law
indices in accelerated populations, highlighting how spectra above 1 MeV
can probe the field obliquity in GRB internal shocks, and the character
of hydromagnetic turbulence in their environs.  It is emphasized that
diffusive shock acceleration theory generates no canonical spectrum at
relativistic MHD discontinuities. This diversity is commensurate with
the significant range of spectral indices discerned in prompt burst
emission. Such system diagnostics are now being enhanced by the
broadband spectral coverage of bursts by the {\it Fermi} Gamma-Ray Space
Telescope; while the Gamma-Ray Burst Monitor (GBM) provides key
diagnostics on the lower energy portions of the particle population, the
focus here is on constraints in the non-thermal, power-law regime of the
particle distribution that are provided by the Large Area Telescope
(LAT).  
\end{abstract}


\end{frontmatter}

\parindent=0.5 cm

\newpage

\section{Introduction}
 \label{sec:Introduction}
Despite rapid evolution of the understanding of gamma-ray bursts over
the last decade since the first redshift determinations established a
cosmological distance scale, there is still much to be learned about
associations, progenitors, and the radiation processes active in both
the prompt and afterglow emission regions.  The most popular burst
paradigm for the genesis of the prompt burst emission is of radiative
dissipation at internal shocks that accelerate particles  (Rees \&
M\'esz\'aros 1992; Piran 1999; M\'esz\'aros 2002).  Within this
scenario, it is of great interest to understand what physical conditions
in the shocked environs can elicit the observed high energy indices and
the spectral structure around the MeV-band peak. It is clear that  the
measurement of the high energy spectral index provides a key constraint
on the interpretation of the electron acceleration process.   To
provides insights into these conditions, one turns to models of
diffusive shock acceleration in relativistic systems.

A core property of acceleration at the relativistic shocks that are
presumed to seed prompt GRB radiation is that the distribution functions
\teq{f(\hbox{\bf p})} are inherently anisotropic. This renders the
power-law indices and other distribution characteristics sensitive to
directional influences such as the magnetic field orientation, and the
nature of MHD turbulence that often propagates along the field lines.
Consequently, familiar results from relativistic shock acceleration
theory such as the so-called canonical \teq{\sigma=2.23} power-law index
(e.g. Kirk et al. 2000) are of fairly limited applicability, though they do
provide useful insights.  This diversity is fortunate since the GRB
database so far has exhibited a substantial range of spectral indices,
so any universal signature in the acceleration predictions would be
unnecessarily limiting.

This paper explores some of the features of diffusive shock acceleration
using results from a test particle Monte Carlo simulation, and addresses
probes of the theoretical parameter space imposed by extant GRB
observations above the MeV spectral break.  The Monte Carlo approach
(Ellison, Jones \& Reynolds 1990; 
Ellison \& Double 2004; Niemiec \& Ostrowski 2004; 
Stecker, et al. 2007) is one of several
major techniques devised to model particle acceleration at shocks;
others include semi-analytic solutions of the diffusion-convection
equation (Kirk \& Heavens 1989; Kirk et al. 2000), and particle-in-cell
(PIC) full plasma simulations (Hoshino, et al. 1992; Nishikawa, et al.
2005; Medvedev, et al. 2005; Spitkovsky 2008). Each has its merits and
limitations.  Tractability of the analytic approaches generally
restricts solution to power-law regimes for the \teq{f(\hbox{\bf p})}
distributions.  PIC codes are rich in their information on shock-layer
electrodynamics and turbulence. To interface with GRB data, a broad
dynamic range in momenta is desirable, and this is the natural niche of
Monte Carlo simulation techniques, the focus of this paper.

Useful diagnostics (Baring \& Braby 2004; Baring 2009) on
\teq{f(\hbox{\bf p})} have already been enabled by data from the BATSE
and EGRET instruments on the Compton Gamma-Ray Observatory (CGRO) for a
few bright bursts. Significant advances are anticipated in the
understanding of such constraints in the next few years, afforded by the
broad spectral coverage and sensitivity of the GBM and LAT experiments
on NASA's {\it Fermi} Gamma-Ray Space Telescope; this is already being
realized.

\section{Diffusive Acceleration at Relativistic Shocks}
 \label{sec:DSA}
The simulation used here to model diffusive acceleration in relativistic
planar shocks is a kinematic Monte Carlo technique that has been
employed extensively in supernova remnant and heliospheric contexts, and
is described in detail in numerous papers (Ellison, Jones \& Reynolds 1990, 
hereafter EJR90; Jones \& Ellison 1991; Ellison, et al. 1995; Ellison \&
Double 2004; Summerlin \& Baring 2006; Baring 2009). It is
conceptually similar to Bell's (1978) test particle approach to
diffusive shock acceleration.  Test particles that are injected upstream
gyrate in laminar electromagnetic fields, their trajectories being
governed by a relativistic Lorentz force equation in the frame of the
shock.  In general, the fluid frame magnetic field is inclined at an
angle \teq{\thetaBone} to the shock normal. Because the shock is moving
with a velocity {\bf u}({\bf x}) relative to the plasma rest frame,
there is, in general, a {\bf u $\times$ B} electric field in addition to
the bulk magnetic field.  Particle interactions with Alfv\'{e}n wave and
other hydromagnetic turbulence are modeled by using a phenomenological
scattering of the charges in the rest frame of the plasma.  The
scattering precipitates spatial diffusion of particles along magnetic
field lines, and to a varying extent, across them as well.  The
scatterings are also assumed to be quasi-elastic, an idealization that
is usually valid because in most astrophysical systems the flow speed
far exceeds the Alfv\'{e}n speed, and contributions from stochastic
second-order Fermi acceleration are small. The diffusion permits a
minority of particles to transit the shock plane numerous times, gaining
energy with each crossing via the shock drift and first-order Fermi
processes.

A continuum of scattering angles, between large-angle or small-angle
cases, can be modeled by the simulation.  In the local fluid frame, the
time, \teq{\delta t_f}, between scatterings is coupled (EJR90) to the 
mean free path, \teq{\lambda}, and the maximum scattering (i.e. 
momentum deflection) angle, \teq{\thetascatt} via \teq{ \delta t_f\approx
\lambda\thetascatt^{2}/(6v)} for particles of speed \teq{v\approx c}. 
Usually \teq{\lambda} is assumed to be proportional to a power of the
particle momentum \teq{p} (see EJR90 and Giacalone, et al.
1992) for microphysical justifications for this choice), and for
simplicity it is presumed to scale as the particle gyroradius,
\teq{r_g}, i.e. \teq{\lambda=\eta r_g\propto p}. The parameter
\teq{\eta} in the model is a measure of the level of turbulence present
in the system, coupling directly to the amount of cross-field diffusion,
such that \teq{\eta =1} corresponds to the isotropic {\it Bohm
diffusion} limit, where the field fluctuations satisfy \teq{\delta
B/B\sim 1}.  In kinetic theory, \teq{\eta} couples the parallel
(\teq{\kappa_{\parallel}=\lambda v/3}) and perpendicular
(\teq{\kappa_{\perp}}) spatial diffusion coefficients via the relation
\teq{\kappa_{\perp}/\kappa_{\parallel}=1/(1+\eta^{2})} (Forman, et al.
1974; Ellison, et al. 1995). In parallel shocks, where the {\bf B} field
is directed along the shock normal (\teq{\thetaBone=0}), \teq{\eta} has
only limited impact on the resulting energy spectrum, principally
determining the diffusive spatial scale normal to the shock.  However,
in oblique relativistic shocks where \teq{\thetaBone > 0}, the diffusive
transport of particles across the field (and hence across the shock)
becomes critical to retention of them in the acceleration process.
Accordingly, for such systems, the interplay between the field angle and
the value of \teq{\eta} controls the spectral index of the particle
distribution (Ellison \& Double 2004; Baring 2004), a feature that is
central to the interpretation of GRB spectra below.

The test particle assumption adopted in this paper is appropriate as
long as the energy density \teq{U_{\rm cr}} of accelerated particles is
much less than that of the thermal gas \teq{U_g}.  This is generally the
case for the simulation results presented here.  For the most energetic
particles to establish \teq{U_{\rm cr}\gtrsim U_g} demands very flat
distributions, namely \teq{\sigma < 2} if \teq{dN/dp = 4\pi p^2
f(\hbox{\bf p})\propto p^{-\sigma}}.  It can be inferred from the
distributions in Fig.~\ref{fig:sas_spec} that those cases that satisfy
this condition are inefficient at injecting from thermal energies, so
that they generate \teq{U_g/U_{\rm cr} \ll 1}.  These contrast
non-linear acceleration scenarios where the non-thermal population
modifies the Rankine-Hugoniot MHD structure of the shock because
\teq{U_{\rm cr}\gtrsim U_g}, thereby inducing spectral concavity in the
energetic tail. Such non-linear cases have been explored extensively for
the contexts of galactic cosmic rays produced in supernova remnant
shells (e.g. Ellison, et al. 2000) and the Earth's bow shock (Ellison,
M\"obius \& Paschmann 1990, hereafter EMP90) 
Generally, as is apparent in
Fig.~\ref{fig:sas_index} below, GRB spectra are too steep to sample the
non-linear acceleration regime. Moreover, in the case of GRB 080916c,
the power-law character above 1 MeV is well established (Abdo, et al.
2009a): no evidence of any spectral concavity can be discerned. While
some models of GRB dissipation in the literature assume that large
fractions (i.e. \teq{\gtrsim 30}\%) of the total lepton energy density
reside in the accelerated electron population (perhaps adopting
truncated power-laws), there is no theoretical mandate for such from the
perspective of diffusive shock acceleration theory.  This assertion
applies not only to results from Monte Carlo codes such as are presented
here, but also to distributions generated in PIC simulations (e.g.
Spitkovsky 2008).

The mildly-relativistic shock regime \teq{u \lesssim c} forms the focus
of this exposition, since this is germane to internal GRB shock models
that invoke shock formation via the collision of two ultra-relativistic
shells.  The discussion will first outline the core characteristics of
the diffusive acceleration process, before moving onto constraints on
theoretical parameters imposed by the observed spectral indices of
prompt emission in bright gamma-ray bursts.

\subsection{Acceleration Characteristics for Relativistic Shocks}
 \label{sec:accel_prop}
As mentioned above, the key property of diffusive acceleration at
relativistic shocks that distinguishes them from their non-relativistic
counterparts is their intrinsic anisotropy.  This is driven by the
powerful convective influence that enables efficient loss of particles
away from and downstream of the shock.  The result of this loss is a
general difficulty in generating flat distributions of shock-accelerated
particles, particularly for so-called superluminal (oblique)
relativistic discontinuities.  Here, the array of possible distribution
indices \teq{\sigma} is highlighted, spawned by the sensitivity of both
the energization in, and escape from, the shock layer, to (i) the size
of the momentum deflection angle \teq{\thetascatt}, (ii) the frequency
or relative mean free path \teq{\lambda/r_g} of scatterings, and (iii)
the upstream field obliquity \teq{\thetaBone}, a quantity derived from
the global MHD structure of the shock.  In this paper, the focus will be
on the latter two influences.

Before investigating them, it is appropriate to mention the first
effect, which was originally identified for relativistic shocks in
EJR90.  When the diffusion in the shock layer samples large field
fluctuations \teq{\delta B/B\sim 1}, it corresponds to large momentum
deflections, delineating the regime of large angle scattering (LAS) with
\teq{4/\Gamma_1\lesssim \thetascatt\lesssim\pi}, where \teq{\Gamma_1} is
the upstream flow's incoming Lorentz factor.  Such large deflections
produce huge gains in particle energy, of the order of \teq{\Gamma_1^2}
in a single scattering and therefore also in successive shock crossings.
These gains are kinematic in origin, and are akin to those in inverse
Compton scattering.  The result is an acceleration distribution
\teq{dN/dp} that is highly structured and much flatter on average than
\teq{p^{-2}}, first noted by EJR90. The bumpy structure is kinematic in
origin and becomes more pronounced for large \teq{\Gamma_1} (Baring
2004; Ellison \& Double 2004; Stecker, et al. 2007, 
hereafter SBS07; Baring 2009).  For ultra-relativistic shocks, when
\teq{p\gg mc}, the bumps asymptotically relax to form a power-law
distribution \teq{dN/dp\propto p^{-\sigma}}, as is mandated by the
concomitant lack of a momentum scale, with an index in the range of
\teq{\sigma\sim 1.6} (SBS07).  From the plasma physics perspective,
magnetic turbulence in relativistic shocks could easily be sufficient 
to effect scatterings on intermediate to large angular scales
\teq{\thetascatt\gtrsim 1/\Gamma_1}, a proposition that becomes more
enticing for ultrarelativistic shocks.

Particle distributions for \teq{\thetascatt\lesssim 1/\Gamma_1} are 
much smoother in appearance, and often necessarily steeper, at least 
for superluminal regimes. Intermediate scattering angles
\teq{\thetascatt\sim 1/\Gamma_1} generate smooth distributions
(SBS07, Baring 2009), much like those for (but flatter than) the more
familiar small angle scattering (SAS, often called pitch angle
diffusion, PAD) which will constitute the regime of focus hereafter. 
This regime has spawned the often cited asymptotic, ultrarelativistic
index of \teq{\sigma =2.23} for \teq{dN/dp\propto p^{-\sigma}}
(Kirk et al. 2000; see also Bednarz \& Ostrowski 1998; Baring 1999).  
This special result is realized only for parallel
shocks with \teq{\thetaBone =0^{\circ}} in the mathematical limit of
small angle scattering \teq{\thetascatt\ll 1/\Gamma_1}, where the
particle momentum is stochastically deflected on arbitrarily small
angular (and therefore temporal) scales.  In such cases, particles
diffuse in the region upstream of the shock only until their velocity's
angle to the shock normal exceeds around \teq{1/\Gamma_1}, after which
they are rapidly swept downstream of the shock.  The lower kinematic
energy gains in shock transits more than compensate for the accompanying
slightly higher shock-layer retention rates, producing a steeper
distribution under SAS conditions.

\begin{figure}
 \centerline{
  \includegraphics[width=.78\textwidth]{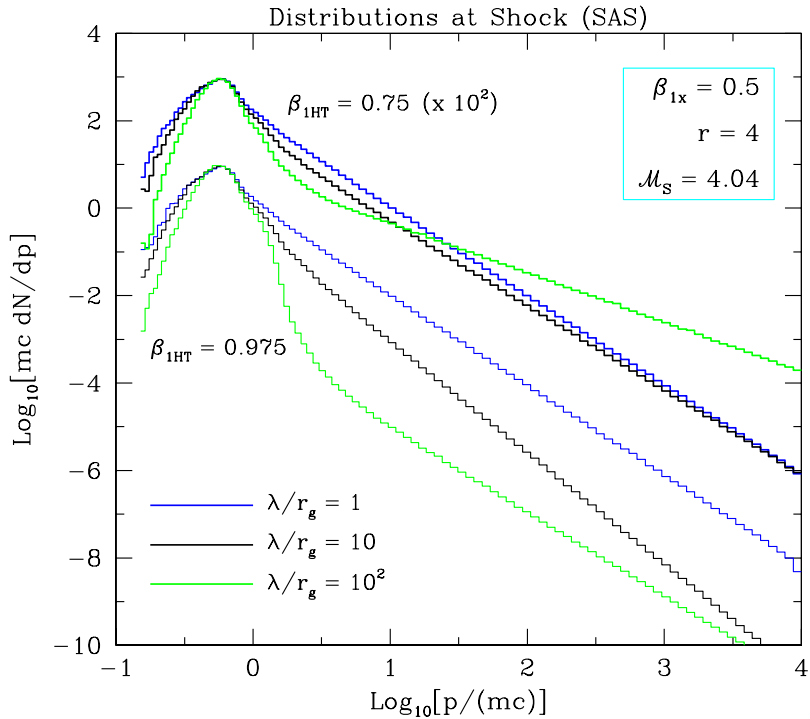}
  }
  \caption{Particle distribution functions  \teq{dN/dp} from mildly-relativistic
sub-luminal shocks (\teq{\Gamma_{1x}\beta_{1x}=0.577}, i.e.
\teq{\beta_{1x}=u_{1x}/c=0.5}) of upstream-to-downstream  velocity
compression ratio \teq{r=u_{1x}/u_{2x}\approx 4}.  Simulation results
are depicted for two upstream fluid frame magnetic field obliquities,
labelled by their corresponding de Hoffman-Teller frame upstream flow
speeds \teq{\beta_{\hbox{\sevenrm 1HT}} = \beta_{1x}/\cos\thetaBone}.
These are in distinct groups of three: \teq{\thetaBone=48.2^{\circ}}
(\teq{\beta_{\hbox{\sevenrm 1HT}} = 0.75}, multiplied by 100) for the
upper three histograms, and \teq{\thetaBone=59.1^{\circ}}
(\teq{\beta_{\hbox{\sevenrm 1HT}} = 0.975}) for the lower three
histograms. Scattering off hydromagnetic turbulence was modeled by
randomly deflecting particle momenta by an angle within a cone, of
half-angle \teq{\thetascatt},  whose axis coincides  with the particle
momentum prior to scattering; three different ratios of the diffusive
mean free path \teq{\lambda} to the gyroradius \teq{r_g} were adopted
for each \teq{\thetaBone}. All results were for small angle scattering
(SAS), when \teq{\thetascatt\lesssim 1/\Gamma_1} and the distributions
become independent of the choice of \teq{\thetascatt}. A low sonic Mach
number \teq{\machson} was chosen so as to maximize the efficiency of
injection from thermal energies.
}
 \label{fig:sas_spec}
\end{figure}

Representative particle differential distributions \teq{dN/dp} that
result from the simulation of diffusive acceleration at
mildly-relativistic (internal GRB) shocks of speed \teq{\beta_{1x}=0.5}
are depicted in Figure~\ref{fig:sas_spec} (see Ellison \& Double 2004,
and SBS07) for \teq{\Gamma_1\gg 1} simulation results).  Here, the
subscript \teq{x} denotes components along the shock normal.  These
distributions are equally applicable to electrons or ions, and were
generated for \teq{\thetascatt \lesssim 10^\circ}, i.e. in the SAS
regime. Results are displayed for two different upstream fluid frame
field obliquities, namely \teq{\thetaBone=48.2^{\circ}} and
\teq{\thetaBone=59.1^{\circ}}. These define shocks with two distinct de
Hoffman-Teller (1950; HT) frame dimensionless speeds \teq{\betaHTone
=\beta_{1x}/\cos\thetaBone}. The HT frame is the shock rest frame where
the flow is everywhere parallel to the local magnetic field.  Note that
the distributions in the Figure were measured in the normal incidence
frame (NIF), the shock rest frame in which the upstream fluid flows in
along the shock normal; in this frame, the magnetic field vectors are
generally not parallel to those in the fluid or HT frames. The HT flow
speed \teq{\betaHTone} corresponds to a physical speed when it is less
than unity, i.e. the upstream field obliquity satisfies
\teq{\cos\thetaBone > \beta_{1x}}; the shock is then called {\it
subluminal}.  When mathematically \teq{\betaHTone > 1}, no Lorentz boost
from the local fluid frame can render the flow parallel to {\bf B} and
de Hoffman-Teller frame does not exist: the shock is said to be {\it
superluminal}.

The distributions clearly exhibit an array of indices \teq{\sigma},
including very flat power-laws, that are not monotonic functions of
either the field obliquity \teq{\thetaBone} nor the key diffusion
parameter \teq{\eta =\lambda /r_g}.  However, it is striking that the
normalization of the power-laws relative to the low momentum thermal
populations is a strongly-declining function of \teq{\lambda /r_g}. 
This is a consequence of a more prolific convection downstream away from
the shock that suppresses diffusive injection from thermal energies into
the acceleration process.  Simulation runs for \teq{\lambda /r_g \geq
10^3} inhibit such injection by several to many orders of magnitude, and
so were not displayed in the plot. Note also that the choice of the
compression ratio \teq{r=4} is somewhat larger than the Rankine-Hugoniot
MHD value for \teq{\beta_{1x}=0.5}, \teq{\machson =4.04} conditions, and
was adopted to afford a direct comparison with the semi-analytic
convection-diffusion equation results of Kirk \& Heavens (1989). Details
of this comparison are discussed briefly in Baring (2009).


\newpage

\begin{figure}
 \centerline{
   \includegraphics[width=.78\textwidth]{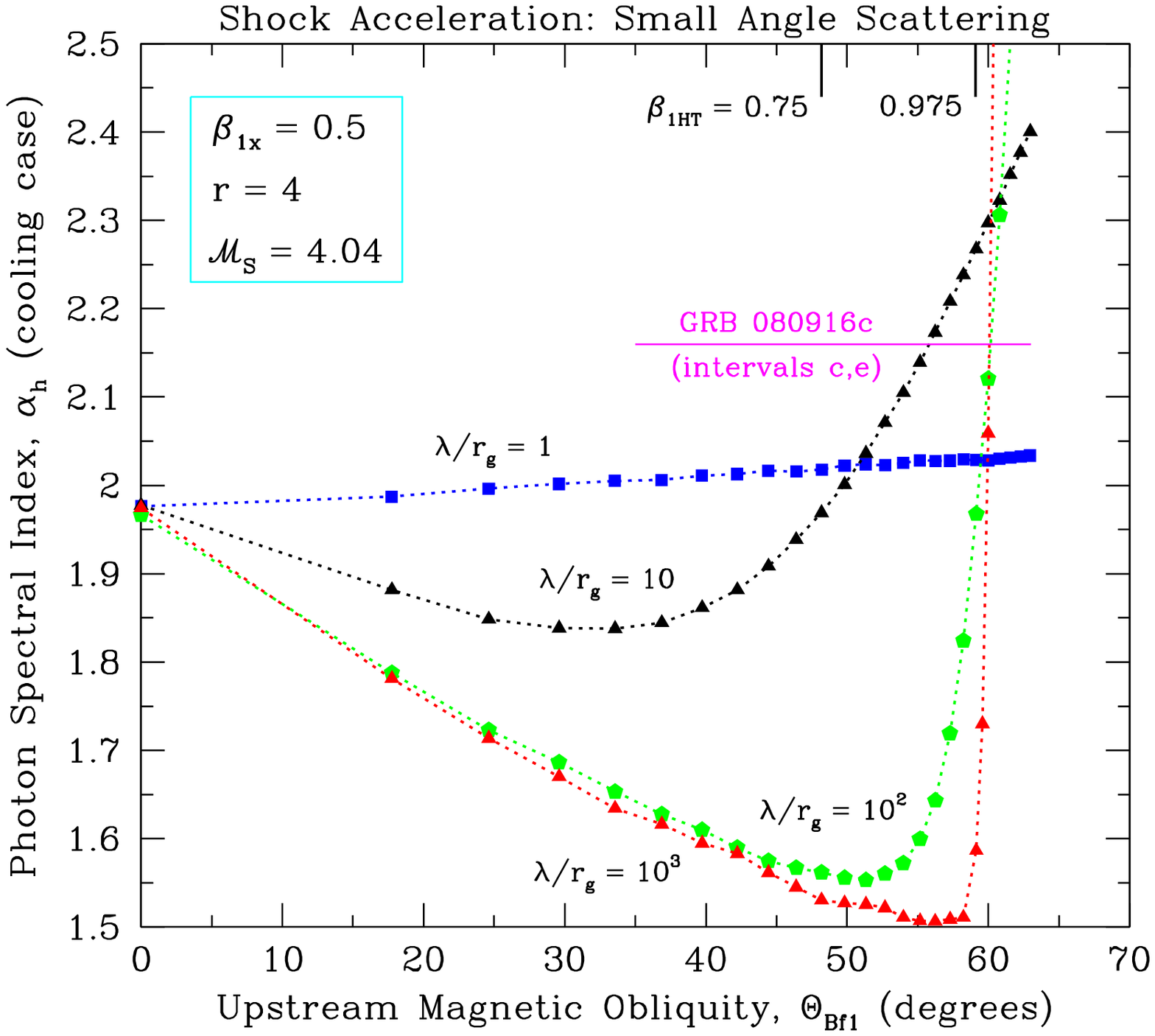}
  }
  \caption{Photon power-law indices \teq{\alpha_h} 
(for \teq{dn/d\erg\propto \erg^{-\alpha_h}}) corresponding to shock
acceleration simulation runs in the limit of small angle scattering,
again for an upstream flow speed \teq{\beta_{1x}\equiv u_{1x}/c =0.5},
and a compression ratio \teq{r=4}. The indices are displayed as
functions of the fluid frame field obliquity \teq{\thetaBone}, with
obliquities \teq{\thetaBone > 60^\circ} constituting superluminal
shocks. The displayed simulation index results were obtained for
different diffusive mean free paths \teq{\lambda} parallel to the mean
field direction, namely \teq{\lambda/r_g=1} (squares),
\teq{\lambda/r_g=10} (triangles), \teq{\lambda/r_g=10^2} (pentagons),
and \teq{\lambda/r_g=10^3} (triangles), as labelled. The heavyweight
horizontal line labelled GRB 080916c indicates the approximate spectral
index \teq{\alpha_h} that is appropriate for this {\it Fermi} burst in
the selected time intervals {\it c} and {\it e} listed in Abdo et al.
(2009a).  All photon indices apply to a cooled synchrotron emission
scenario: see the text for a discussion.
}
 \label{fig:sas_index}
\end{figure}

To summarize the influence of the two key parameters that dictate the
range of possible spectral indices \teq{\sigma}, namely the field
obliquity \teq{\thetaBone}, and the ratio \teq{\eta =\lambda /r_g},  a
parameter survey for diffusive acceleration at \teq{\beta_{1x}} shocks
(typical of mildly-relativistic systems) is exhibited in
Figure~\ref{fig:sas_index}. Again, the small angle scattering limit was
employed. To interface more directly with GRB observations, it is
convenient to represent the particle power-law indices via their
radiative emission equivalent, namely the high energy spectral index
\teq{\alpha_h} (or the Band model index \teq{\beta = -\alpha_h} of
common usage: see Band et al. 1993), for which the observed differential
photon spectrum is \teq{dn/d\erg\propto \erg^{-\alpha_h}} above the
MeV-band \teq{\nu}-\teq{F_{\nu}} peak. The correspondence between
\teq{\sigma} and \teq{\alpha_h} depends on the emission mechanism and
the model assumptions.  Here, the popular quasi-isotropic synchrotron
mechanism (Rees \& M\'esz\'aros 1992; Tavani 1996; Piran 1999;
M\'esz\'aros 2002) in bursts is adopted for this purpose.  Moreover, for
this illustrative agenda, it is presumed that the emission is in the
strongly-cooling domain, for which an accelerated electron/pair
distribution with index \teq{\sigma} steepens under synchrotron cooling
to a power-law of index of \teq{\sigma +1}.  With these specifications,
one simply has the relation \teq{\alpha_h = (\sigma + 2)/2}, so that
\teq{p^{-2}} shock acceleration power-laws map over to \teq{\erg^{-2}}
photon spectra.  Accordingly, the photon power-law index \teq{\alpha_h}
is plotted as a function of \teq{\thetaBone}, with subluminal shocks
constituting those with obliquities \teq{\thetaBone < 60^\circ}. There
is clearly a considerable range of possible photon indices spawned by
non-thermal particles accelerated in mildly relativistic shocks --- the
same is true for hadronic emission components, where, quite often
\teq{\alpha_h \approx \sigma}).  This is a non-universality that can
attractively mesh with GRB observations.  Note also that the
distributions in Fig.~\ref{fig:sas_spec} correspond to indices at the
obliquities marked in Fig.~\ref{fig:sas_index} at the top by the labels
\teq{\beta_{\hbox{\sevenrm 1HT}} = 0.75} and 
\teq{\beta_{\hbox{\sevenrm 1HT}} = 0.975}.


An obvious feature of this plot is that the dependence of \teq{\alpha_h}
on field obliquity is non-monotonic.  When \teq{\lambda /r_g\gg 1}, the
value of \teq{\alpha_h} at first declines as \teq{\thetaBone} increases
above zero. This leads to very flat spectra.  As \teq{\betaHTone}
approaches and eventually exceeds unity, this trend reverses, and
\teq{\alpha_h} then rapidly increases with increasing shock obliquity. 
This dramatic steepening of the distribution in near-luminal and
superluminal shocks is precipitated by inexorable convection of
particles away downstream of the shock, character that is evinced in
Fig.~\ref{fig:sas_spec}.  Any amelioration of this rapid decline in the
acceleration efficiency requires the reduction of \teq{\lambda /r_g} to
values below around \teq{10}.  Physically, this condition is tantamount
to increasing the hydromagnetic turbulence to high levels that force the
particle diffusion to approach isotropy, i.e. encroaching upon the {\it
Bohm diffusion} limit of \teq{\lambda /r_g\sim 1}.  Then, transport of
charges across the mean field becomes significant on gyrational
timescales, and they can be retained near the shock for sufficient times
to accelerate and generate suitably flat distribution functions.  Such
low values of \teq{\lambda /r_g} render the field direction immaterial,
and the shock behaves much like a parallel, subluminal shock in terms of
its diffusive character. Then, \teq{\alpha_h} is only weakly dependent
on \teq{\thetaBone}, another important property illustrated in 
Fig.~\ref{fig:sas_index}.

The third key characteristic is the very flat spectra with
\teq{\alpha_h\sim 1.5} that form from extremely flat electron
distributions with \teq{\sigma\sim 1}. The origin of these (Baring 2009)
is in the coherent effect of {\it shock drift acceleration} at the shock
discontinuity.  This phenomenon is due to the energy gain of charges
when they repeatedly encounter {\bf u}\teq{\times}{\bf B} drift electric
fields (in frames other than the HT frame) in gyrations straddling the
shock discontinuity; it has been widely discussed in the context of
non-relativistic astrophysical and heliospheric shocks.  When
\teq{\lambda /r_g\gg 1}, the charge trajectories maintain helical
coherence so that for select gyrophases they can be trapped in the shock
layer, gain energy and be reflected upstream to subsequently participate
in repeated shock encounters.  Particles can then efficiently gain
energy by the shock drift process before they are eventually lost
downstream, and the result is a distribution function approaching
\teq{dN/dp \propto p^{-1}}.   Reducing \teq{\lambda /r_g} and thereby
introducing extremely modest amounts of turbulence and associated
cross-field diffusion disrupts this coherence, removes particles from
the shock layer, and accordingly steepens the spectrum.

\section{GRB Observations as Probes of Shock Acceleration}
\label{sec:obs_vs_theory}

The shock acceleration theory results presented in
Figs.~\ref{fig:sas_spec} and~\ref{fig:sas_index} can now be interpreted
in the light of prompt GRB observations.  The focus here is on the high
energy spectral index \teq{\alpha_h} of the power-law above the
\teq{\nu}-\teq{F_{\nu}} peak.  This makes direct connection to data from
CGRO's EGRET telescope, and now to the growing database of {\it Fermi}
LAT burst detections.  To complement these probes of diffusive shock
acceleration modeling of gamma-ray bursts, it must be noted that
considerable insights can also be gleaned from consideration of the
burst spectral shape at and below the \teq{\nu}-\teq{F_{\nu}} peak. This
was the principal focus of the investigation by Baring \& Braby (2004),
that built upon the early work on broad-band spectral fitting of bursts
was provided by Tavani (1996).  The constraints on particle
distributions, shock acceleration interpretations and viable radiation
mechanisms in bright CGRO bursts that were derived in Baring \& Braby
(2004) are reviewed in Baring (2009).

It should be noted that forging a direct connection between the observed
\teq{\alpha_h} and the underlying non-thermal particle distribution
index \teq{\sigma} is possible when the Thomson optical depth
\teq{\taut} is small, and is straightforward for most bursts with good
spectroscopy and well-established power-laws above 1 MeV.  The {\it
Fermi}-LAT bursts GRB 080916c (Abdo, et al. 2009a) and GRB 090510 (Abdo
et al. 2009b) are suitable examples.  Occasionally, the situation may be
more complicated.  For example, some models adopt high \teq{\taut} (e.g
M\'esz\'aros \& Rees 2000; Pe'er \& Waxman 2004; Ryde 2005) that lead to
strong Comptonization/thermalization and high compactnesses that spawn
rampant pair creation. These can prove viable for isolated instances
when the observed spectrum is quasi-thermal (e.g. Ryde 2005), and a jet
photosphere is invoked. A case in point is time interval {\it b} in the 
{\it Fermi}-LAT burst GRB 090902b (Abdo et al. 2009c), which offers a 
highly-peaked and narrow spectral component that may (or may not) 
preclude a shock acceleration interpretation.  Yet, this is an exceptional 
selection, and most GRB observations offer fairly clean \teq{\alpha_h<3.5} 
determinations that permit inferences of an underlying electron or proton 
{\it non-thermal distribution} index \teq{\sigma}.  In the majority of extant 
burst detections, the non-thermal spectroscopic interpretation still remains 
a simpler explanation than does the non-isothermal convolution of spectra
close to the ideal Planck form.

The most extensive database for high-energy photon spectral indices
\teq{\alpha_h} in GRBs is for the CGRO EGRET and Comptel bursts.  The
EGRET \teq{\alpha_h} index distribution (e.g. Dingus 1995) is constituted
by a handful of sources with indices scattered in the range
\teq{2\lesssim\alpha_h\lesssim 3.7}. Of these, EGRET spark chamber
detections were concentrated in the index range \teq{\alpha_h \lesssim
2.8}, as tabulated in Baring (2006).  One of the flattest EGRET burst
spectra was for GRB 930131, with an index \teq{\alpha_h \approx 2},
which corresponds to the indication of \teq{\lambda /r_g\lesssim 10}
from Fig.~\ref{fig:sas_index} (or almost luminal, but not superluminal
shocks) if a synchrotron or inverse Compton cooling model is adopted. 
The CGRO BATSE  \teq{\alpha_h} index distribution (Preece, et al. 2000) is
similarly broad, but with \teq{1.5\lesssim \alpha_h\lesssim 3.5} and far
greater statistics.   The recent {\it Fermi} detection (Abdo, et al. 2009a) of
GRB 080916c in both the GBM and LAT instruments offered a high energy
index of \teq{\alpha_h\sim 2.2} (marked in Fig.~\ref{fig:sas_index}) in
its most luminous epochs, and a generally steeper spectrum at other
times. Accordingly, observationally, shock acceleration models must
accommodate a radiation spectral index in the range
\teq{2\lesssim\alpha_h\lesssim 4} in order to be viable.  Moreover, they
must reasonably account for the spectral variability identified in GRB
080916c, i.e. fluctuating \teq{\alpha_h} values.  This connects to a new
development afforded by the refined sensitivity for spectroscopy of {\it
Fermi}: spectral evolutionary studies in temporal sub-intervals are now
a reality, providing a boon for model diagnostics and a honing of the
burst paradigm.

Fig.~\ref{fig:sas_index} displays photon indices for electron
synchrotron emission in strongly cooling scenarios following shock
acceleration, where \teq{\alpha_h = (\sigma + 2)/2}.  It is clear that
the flattest spectral epochs for GRB 080916c are best described by
highly-oblique, mildly-relativisic shocks, but not quasi-perpendicular
(\teq{\thetaBone\gtrsim 70^\circ}) superluminal ones.  This can also be
inferred for the time-integrated spectral behavior of the EGRET burst
GRB 910503 (see Baring, 2006). In contrast, flat-spectrum EGRET
bursts GRB 930131 and GRB 950425 with \teq{\alpha_h\lesssim 2} cannot
correspond to superluminal acceleration regimes if radiative cooling is
rampant, but demand subluminal shocks with moderate to strong turbulence
therein, i.e. \teq{\lambda /r_g\sim 3 - 10}. Other EGRET bursts such as
GRB 910601 and GRB 990123 (again see Baring 2006 for tabulated
indices), and some sub-intervals in the {\it Fermi} dataset
(Abdo, et al. 2009a) for GRB 080916c exhibit indices \teq{\alpha_h > 2.5}. 
From Fig.~\ref{fig:sas_index}, these bursts can be consistent with
acceleration at mildly superluminal shocks in the
\teq{60^\circ}-\teq{70^\circ} obliquity range, but only if the
scattering is strong, i.e. \teq{\lambda /r_g\lesssim 3-10}.  No bursts
have so far evinced extended power-law spectra flatter than
\teq{\alpha_h\approx 1.95} above the the \teq{\nu}-\teq{F_{\nu}} peak
(the possibility of additional high energy components above 100 MeV in
GRB 940217 [Hurley, et al. 1994], GRB 941017 [Gonzalez, et al. 2003]
and more recently GRB 090902b [Abdo, et al. 2009c], is a separate
issue), absolving the need for acceleration in shocks with extremely low
turbulence, i.e. \teq{\lambda/r_g\gtrsim 10^2} regimes.  This is
fortunate, since such shocks are inherently inefficient accelerators.
Moreover, the generation of field turbulence is a natural part of
dissipation in shocks, so that nearly laminar fields are not expected. 
Field structures that are devoid of fluctuations are never observed in
{\it in situ} magnetometer measurements at heliospheric shocks (e.g.
Baring et al. 1997, and references therein). Note that moderate field
obliquities are quite plausible in mildly-relativistic internal shocks,
contrasting ultra-relativistic external GRB shocks, which are
necessarily quasi-perpendicular (\teq{\thetaBone\approx 90^{\circ}}) due
to the Lorentz transformation of circumburst fields.

\newpage

These inferences are predicated on the presumption of efficient cooling
in the burst prompt emission zone.  This must operate over a significant
range of electron momenta, since cooling breaks, by an index of
\teq{\Delta\alpha_h =1/2}, are not observed above 1 MeV in burst
emission.  For example,  if the spectra spanning the range 1 MeV -- 10
GeV in GRB 080916c correspond to a cooling-dominated contribution, the
acceleration must operate in short, impulsive periods, followed by long
cooling epochs that permit the electrons to decline in momentum by a
factor of \teq{10^2} or so.  Then the cooling epochs should possess
durations of around 4 orders of magnitude longer than the impulsive
acceleration epochs.  This is a significant constraint on cooling
scenarios within the internal shock model (Baring 2009). The alternative
of uncooled synchrotron emission yields a photon differential spectral
index given by \teq{\alpha_h = (\sigma + 1)/2} above the
\teq{\nu}-\teq{F_{\nu}}  peak.  Then the injected distribution must have
an index \teq{\sigma} higher than that for cooling models by unity, in
order to match the burst observations. Specifically, the EGRET GRB index
range \teq{2\lesssim\alpha_h\lesssim 3.7} maps over to
\teq{3\lesssim\sigma\lesssim 6.4}. This is a profound difference in that
it pushes the viable shock parameter space into the superluminal range,
i.e. at higher field obliquities, and Bohm-limited diffusion is
observationally excluded. In addition, for the less popular uncooled
hadronic models, the photon and particle indices trace each other, i.e.
\teq{\alpha_h = \sigma}, and a similar conclusion that the environment
is restricted to modestly superluminal oblique shocks is derived.

It must be emphasized that the spectral index array in
Fig.~\ref{fig:sas_index} is quite representative of the range and
parametric behavior for general mildly-relativistic shocks.  While the
subluminal/superluminal obliquity boundary varies with \teq{\beta_{1x}},
the morphology of the index curves, their trends with \teq{\lambda/r_g},
and their rapid transition to large indices when crossing to the
superluminal domain remain qualitatively the same for shock speeds in
the range \teq{0.1\lesssim\beta_{1x}\lesssim 0.9}.

\section{Conclusion}

This paper has explored constraints imposed on the parameter space for
diffusive acceleration at relativistic shocks by prompt emission
observations in gamma-ray bursts.  The simulation results presented
showcase the non-universality of the index of non-thermal particles,
spawned by a range of shock obliquities and the varied character of
hydromagnetic turbulence in their environs.  This non-universality poses
no problem for modeling GRB high-energy power-law indices
\teq{\alpha_h}.  The observations generally constrain the shock
parameter space to oblique, subluminal or highly-turbulent superluminal
shocks not far the Bohm diffusion limit, i.e. \teq{\lambda /r_g\lesssim
10}. This presumes rapidly cooling synchrotron or inverse Compton
emission scenarios. If cooling arises on timescales exceeding those
pertinent for acceleration, then the acceleration must occur in mildly
superluminal, turbulent shocks; subluminal shocks produce particle
distributions too flat to accommodate the observed spectral indices.  
The high photon count detections of the long duration GRB 080916c and 
the short burst GRB 090510 in both {\it Fermi's} GBM and LAT
telescopes has afforded a new opportunity to constrain shock
acceleration parameters in temporal sub-intervals.  The prospect of a
number of similar platinum standard, broad-band GRB detections by 
{\it Fermi} that permit time-dependent spectroscopy should hone our
understanding of the connection between shock acceleration and prompt
emission.
\vskip 10pt
\noindent
{\bf Acknowledgments:} this research was supported in part by National 
Science Foundation grant PHY07-58158 and NASA grant NNG05GD42G.  
I am also grateful to the Kavli Institute for Theoretical Physics, University
of California, Santa Barbara for hospitality during part of the period when 
this research was performed, a visit that  was supported in part by the 
National Science Foundation under Grant No. PHY05-51164.

\section{References}


\setlength{\parskip}{.00in}



\reference 
Abdo, A.~A., Ackerman, M., Arimoto, M., et al. ({\it Fermi} LAT Collaboration) 2009a, 
Science, \vol{323}{1688--1693}
``{\it Fermi} Observations of High-Energy Gamma-Ray Emission from GRB 080916c."
\reference 
Abdo, A.~A., Ackerman, M., Ajello, M., et al. ({\it Fermi} LAT Collaboration) 2009b, 
Nature, \vol{462}{331--334}
``{\it Fermi} Gamma-Ray Burst GRB090510 Observations 
    Limit Variation of Speed of Light with Energy."
\reference 
Abdo, A.~A., Ackerman, M., Ajello, M., et al. ({\it Fermi} LAT Collaboration) 2009c, 
\apj, \vol{706}{L138--L144}
``{\it Fermi} Observations of GRB 090902B: A Distinct Spectral 
     Component in the Prompt and Delayed Emission."
\reference 
Band, D.~L., Matteson, J., Ford, L., et al. 1993, \apj,\vol{413}{281--292}
``BATSE Observations of Gamma-Ray Burst Spectra. I - Spectral Diversity."
\reference 
Baring, M. G.  1999, in \it Proc. of the 26th ICRC, Vol. IV \rm ,
     pp.~5--8,
``Acceleration at Relativistic Shocks in Gamma-Ray Bursts"  {\tt [astro-ph/9910128]}.
\reference 
Baring, M.~G. 2004, Nucl. Phys. B, \vol{136C}{198--207}
``Diffusive Shock Acceleration of High Energy Cosmic Rays."
\reference 
Baring, M.~G.  2006, \apj,\vol{650}{1004--1019}
``Temporal Evolution of Pair Attenuation Signatures in Gamma-Ray Burst Spectra."
\reference 
Baring, M.~G. 2009, in {\it  Proc. 6th Huntsville GRB Symposium},
     eds. C.~A. Meegan, et al., N. Gehrels, \& C. Kouveliotou
     (AIP Conf. Proc. 1133, New York) p.~294--299. {\tt [astro-ph/0901.2535]}
``Probes of Diffusive Shock Acceleration using Gamma-Ray Burst Prompt Emission."
\reference 
Baring, M.~G.  \& Braby, M.~L. 2004, \apj,\vol{613}{460--476}
``A Study of Prompt Emission Mechanisms in Gamma-Ray Bursts." (BB04)
\newpage
\reference 
Baring, M. G., Ogilvie, K. W., Ellison, D., \& Forsyth, R. 1997, 
   \apj,\vol{476}{889--902}
``Acceleration of Solar Wind Ions by Nearby Interplanetary Shocks:
      Comparison of Monte Carlo Simulations with Ulysses Observations."
\reference 
Bednarz, J. \& Ostrowski, M. 1998, \prl\vol{80}{3911--3914}
``Energy Spectra of Cosmic Rays Accelerated at Ultrarelativistic Shock Waves."
\reference 
Bell, A.~R. 1978, \mnras\vol{182}{147--156}
``The Acceleration of Cosmic Rays in Shock Fronts. I."
\reference 
de Hoffman, F. \& Teller, E. 1950, \prd,\vol{80}{692--703}
``Magneto-Hydrodynamic Shocks."
\reference 
Dingus, B.~L. 1995, \apss,\vol{231}{187--190}
``EGRET Observations of $> 30$ MeV Emission from the Brightest Bursts Detected by BATSE."
\reference 
Ellison, D.~C., Baring, M.~G. \& Jones, F.~C. 1995, \apj,\vol{453}{873--882}
``Acceleration Rates and Injection Efficiencies in Oblique Shocks."
\reference 
Ellison, D.~C., Berezhko, E.~G. \& Baring, M.~G. 2000, \apj\vol{540}{292--307}
``Nonlinear Shock Acceleration 
   and Photon Emission in Supernova Remnants."
\reference 
Ellison, D.~C. \& Double, G.~P. 2004, \app,\vol{22}{323--338}
``Diffusive Shock Acceleration in Unmodified Relativistic, Oblique Shocks."
\reference 
Ellison, D.~C., Jones, F.~C. \& Reynolds, S.~P. 1990, \apj,\vol{360}{702--714}
``First-Order Fermi Particle Acceleration by Relativistic Shocks." (EJR90)
\reference 
Ellison, D.~C., M\"obius, E., \& Paschmann, G. 1990, \apj\vol{352}{376--394}
``PArticle Injection and Acceleration at Earth's Bow Shock 
-- Comparison of Upstream and Downstream Events." (EMP90)
\reference 
Forman, M.~A., Jokipii, J.~R. \& Owens, A.~J. 1974, \apj,\vol{192}{535--540}
``Cosmic-Ray Streaming Perpendicular to the Mean Magnetic Field."
\reference 
Giacalone, J., Burgess, D., \& Schwartz, S. J. 1992, in 
   {\it ESA, Study of the Solar-Terrestrial System}, p.~65--70.
   ``Ion Acceleration at Parallel Shocks: Self-Consistent Plasma Simulations."
\reference 
Gonz\'alez, M.~M., Dingus, B.~L., Kaneko, Y., et al. 2003, 
   Nature, \vol{424}{749--751} ``A Gamma-Ray Burst with a High-Energy Spectral 
   Component Inconsistent with the Synchrotron Shock Model."
\reference 
Hoshino, M., Arons, J.,  Gallant, Y.~A. \& Langdon, A.~B. 1992 \apj,\vol{390}{454--479}
``Relativistic Magnetosonic Shock Waves in Synchrotron Sources 
-- Shock Structure and Nonthermal Acceleration of Positrons."
\reference 
Hurley, K., Dingus, B. L., Mukherjee, R., et al. 1994, \nat,\vol{372}{652--652}
``Detection of a Gamma-Ray Burst of Very Long Duration and Very High Energy."
\reference 
Jones, F.~C. \& Ellison, D.~C. 1991, \ssr,\vol{58}{259--346}
``The Plasma Physics of Shock Acceleration."
\reference 
Kirk, J.~G., Guthmann, A.~W., Gallant, Y.~A., Achterberg, A. 2000, \apj,\vol{542}{235--242}
``Particle Acceleration at Ultrarelativistic Shocks: An Eigenfunction Method."
\reference 
Kirk, J.~G. \& Heavens, A.~F. 1989, \mnras,\vol{239}{995--1011}
``Particle Acceleration at Oblique Shock Fronts."
\newpage
\reference 
Medvedev, M.~V., Fiore, M., Fonseca, R.~A., et al. 2005, \apj,\vol{618}{L75--L78}
``Long-Time Evolution of Magnetic Fields in Relativistic Gamma-Ray Burst Shocks."
\reference 
M\'esz\'aros, P. 2002, Ann. Rev. Astron. Astr., \vol{40}{137--169}
``Theories of Gamma-Ray Bursts."
\reference 
M\'esz\'aros, P. \& Rees, M.~J. 2000, \apj, \vol{530}{292--298}
``Steep Slopes and Preferred Breaks in Gamma-Ray Burst Spectra: 
   The Role of Photospheres and Comptonization."
\reference 
Niemiec, J., \& Ostrowski, M. 2004, \apj,\vol{610}{851--867}
``Cosmic-Ray Acceleration at Relativistic Shock Waves with a ``Realistic'' Magnetic Field Structure"
\reference 
Nishikawa, K.-I., Hardee, P., Richardson, G., et al. 2005, \apj,\vol{622}{927--937}
``Particle Acceleration and Magnetic Field Generation in Electron-Positron Relativistic Shocks."
\reference 
Pe'er, A. \& Waxman, E. 2004, \apj,\vol{613}{448--459}
``Prompt Gamma-Ray Burst Spectra: Detailed Calculations and the Effect of Pair Production."
\reference 
Piran, T.  1999, Phys. Rep., \vol{314}{575--667}
``Gamma-Ray Bursts and the Fireball Model."
\reference 
Preece, R.~D., Briggs, M.~S., Mallozzi, R.~S., et al. 2000, \apj\ Supp.,\vol{126}{19--36}
``The BATSE Gamma-Ray Burst Spectral Catalog. 
   I. High Time Resolution Spectroscopy of Bright Bursts Using High Energy Resolution Data."
\reference 
Rees, M.~J. \& M\'esz\'aros, P. 1992, \mnras,\vol{258}{41P--43P}
``Relativistic Fireballs - Energy Conversion and Time-Scales."
\reference 
Ryde, F. 2005, \apj, \vol{625}{L95--L98}
``Is Thermal Emission in Gamma-Ray Bursts Ubiquitous?"
\reference 
Spitkovsky, A. 2008, \apj,\vol{682}{L5--L8}
``Particle Acceleration in Relativistic Collisionless Shocks: Fermi Process at Last?"
\reference 
Stecker, F.~W., Baring, M.~G. \& Summerlin, E.~J. 2007, \apj,\vol{667}{L29--L32}
``Blazar Gamma-Rays, Shock Acceleration, and the Extragalactic Background Light." (SBS07)
\reference 
Summerlin, E.~J. \& Baring, M.~G. 2006, \asr,\vol{37(8)}{1426--1432}
``Modeling Accelerated Pick-up Ion Distributions at an Interplanetary Shock."
\reference 
Tavani, M. 1996, \prl,\vol{76}{3478--3481}
``Shock Emission Model for Gamma-Ray Bursts."


\end{document}